\documentclass[sigconf]{acmart}
\usepackage[utf8]{inputenc}
\usepackage[labelfont=bf]{caption}
\usepackage[justification=centering]{caption}
\usepackage{esvect}
\usepackage{amsmath}

\date{April 2020}

\usepackage{graphicx}
\setcopyright{acmcopyright}
\copyrightyear{2020} 
\acmYear{2020} 
\setcopyright{rightsretained} 
\acmConference[ASIA CCS '20]{Proceedings of the 15th ACM Asia Conference on Computer and Communications Security}{October 5--9, 2020}{Taipei, Taiwan}
\acmBooktitle{Proceedings of the 15th ACM Asia Conference on Computer and Communications Security (ASIA CCS '20), October 5--9, 2020, Taipei, Taiwan}\acmDOI{10.1145/3320269.3405438}
\acmISBN{978-1-4503-6750-9/20/10}
\begin{document}
\fancyhead{}

\title{Attacks to Federated Learning: Responsive Web User Interface to Recover Training Data from User Gradients}

\author[1]{Hans Albert Lianto}
\affiliation{
\institution{Nanyang Technological University}
\country{Singapore}}
\email{HANS0032@e.ntu.edu.sg}
\authornote{Both authors contributed equally to the paper. The order of names is alphabetical.}

\author[2]{Yang Zhao}
\affiliation{
\institution{Nanyang Technological University}
\country{Singapore}}
\email{S180049@e.ntu.edu.sg}
\authornotemark[1]
\authornote{Corresponding author}

\author[3]{Jun Zhao}
\affiliation{
\institution{Nanyang Technological University}
\country{Singapore}}
\email{junzhao@ntu.edu.sg}

\begin{abstract}
Local differential privacy (LDP) is an emerging privacy standard to protect individual user data. One scenario where LDP can be applied is federated learning, where each user sends his/her user gradients to an aggregator who uses these gradients to perform stochastic gradient descent. In a case where the aggregator is \mbox{untrusted} and LDP is not applied to each user gradient, the aggregator can recover sensitive user data from these gradients. In this paper, we present an interactive web demo showcasing the power of LDP by visualizing federated learning with LDP. Moreover, the live demo shows how LDP can prevent untrusted aggregators from recovering sensitive training data. A measure called the \textit{exp-hamming recovery} is also created to show the extent of how much data the aggregator can recover.
\end{abstract}

\ccsdesc[500]{Security and privacy~Distributed systems security}
\ccsdesc[300]{Computing methodologies~Supervised learning by classification}
\ccsdesc[300]{Computing methodologies~Supervised learning by regression}

\keywords{Local differential privacy, Machine learning, Federated learning, Linear regression, Stochastic gradient decent.}

\maketitle
 \thispagestyle{fancy}
\pagestyle{fancy}
\lhead{This paper appears in ACM ASIA Conference on Computer and Communications Security (ACM \textbf{ASIACCS}) 2020.\\ Please feel free to contact us for questions or remarks.}
\cfoot{\thepage}
\renewcommand{\headrulewidth}{0.4pt}
\renewcommand{\footrulewidth}{0pt}

\section{Introduction}
\subsection{Local differential privacy}
Local differential privacy (LDP) is used to perturb data locally. Even when perturbed data are exposed to adversaries (which could include the data curator or aggregator), LDP guarantees that  selected user’s sensitive information is not leaked. Hence even the data curator or aggregator is not trusted with true information from each entry in the dataset. Local differential privacy was implemented by Google in its RAPPOR technology in their paper in \cite{erlingsson2014rappor}.

Local differential privacy ($\epsilon$-LDP) \cite{erlingsson2014rappor} is defined as follows:

\begin{definition}
A randomized algorithm $f$ satisfies $\epsilon$-local differential privacy if and only if, for any two tuples $t$ and $t'$ in the domain of $f$, and for all subsets S of the output range:
\[Pr[f(t) \in S] \leq e^\epsilon \times Pr[f(t') \in S],\]
where parameter $\epsilon$ is the privacy budget and $Pr[\cdot ]$ is probability.
\end{definition}

Many companies store enterprise data in relational form with RDBMS software. For example, Airbnb and Uber store their data using mySQL \cite{stackshare}; Netflix and Instagram store their data using PostgreSQL \cite{stacksharep}. Data stored in this relational form come in a set of tuples or records with the same schema. Each tuple or record have the same number of ‘dimensions’ or rows. LDP-perturbed data can hence also be stored in these relational forms.

Methods to perturb this multidimensional data to satisfy $\epsilon$-LDP have been proposed, including Duchi \emph{et~al.}’s recent proposal in \cite{duchi2018minimax} and Wang’s improved proposal (the Piecewise Mechanism and Hybrid Mechanism) in \cite{wang2019collecting}. These methods will be used in the web demo outlined in subsequent sections of the paper.

\subsection{Local differential privacy applied in federated learning}
\begin{figure}[h!]
\centering
\includegraphics[scale=0.7]{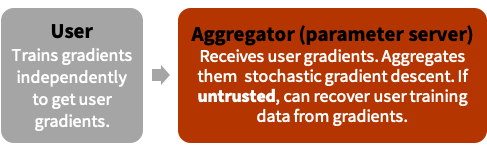}\vspace{-10pt}
\caption{Federated learning flowchart}\vspace{-10pt}
\captionsetup{justification=centering}
\label{fig:federated}
\end{figure}

One practical application for local differential privacy is in machine learning algorithms, particularly in federated or centralized distributed learning. As illustrated in Fig. \ref{fig:federated}, each user is considered a node that trains gradients independently; each node then sends these gradients to the parameter server for aggregation \cite{zhu2019deep}. The data that these users make use of to obtain user gradients is sensitive information and should not be leaked.

\vspace{-5pt}

It can be shown in the subsequent web demo that without local differential privacy, a significant portion of the training data could be leaked by the untrusted aggregator in any federated learning setting where each user submits their gradients for training.

\section{Demo Overview}

\begin{figure}[h!]
\centering
\frame{\includegraphics[scale=1]{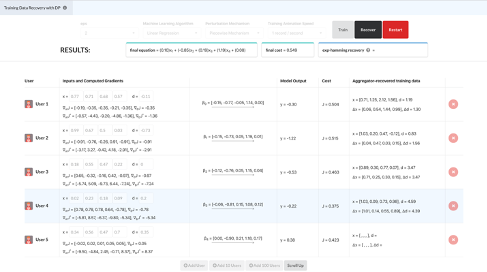}}\vspace{-10pt}
\caption{Screenshot of ldp-machine-learning demo}\vspace{-10pt}
\captionsetup{justification=centering}
\label{fig:ldpmachinelearning}
\end{figure}

A web user interface (called ldp-machine-learning) is created to simulate stochastic gradient descent in federated learning, and for the aggregator to recover training data from each user. A screenshot of the demo is shown in Fig. \ref{fig:ldpmachinelearning} and the demo is publicly available in \url{https://ldp-machine-learning.herokuapp.com/} 

In the demo, when all users fill in the training data and the Train button is started, one epoch of stochastic gradient descent takes place. Unbeknownst to the ‘users’ illustrated in the GUI, the aggregator uses the gradients from the user and the current parameters of the model to backtrack and recover the training user data, which poses a security risk. Features of the UI are outlined in the next section.

\subsection{Animation and animation speed}
The main animation in the demo occurs when the Train and Recover buttons are clicked. When the Train button is clicked, training proceeds and the current cost or training accuracy of the machine learning model is shown beside each user sending his or her gradients to the aggregator. When training is finished, the final model cost and accuracy are outputted on screen. When the Recover button is clicked, the untrusted aggregator attempts to recover sensitive user training data from the gradients sent by each user. At the end of recovery, the average \textit{exp-hamming recovery} for a user is outputted on screen. Normally, the training and recovery animation occurs at a rate of one user per second; however, changing the Training Animation Speed input from “1 record / second” to “Instant” in the UI input form will remove the animation and immediately display training and recovery results when training finishes.

\subsection{Machine learning algorithms and LDP perturbation mechanisms}
\begin{figure}[h!]
\centering
\includegraphics[scale=0.7]{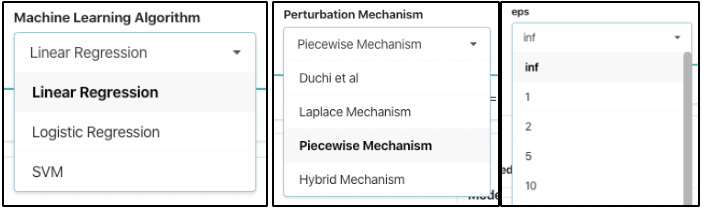}
\caption{Toggling between different ML algorithms, LDP algorithms and privacy budgets}
\captionsetup{justification=centering}\vspace{-10pt}
\label{fig:inputtoggle}
\end{figure}

The demo features federated learning with various machine learning algorithms such as linear regression, logistic regression and support vector machine, which can be toggled, as shown in Fig. \ref{fig:inputtoggle}.

Moreover, as shown in Fig. \ref{fig:inputtoggle}, the LDP perturbation mechanism to perturb user gradients can be toggled from 4 options: Laplace mechanism, Duchi~\emph{et~al.}'s mechanism~\cite{duchi2018minimax}, Piecewise mechanism~\cite{wang2019collecting} and Hybrid mechanism~\cite{wang2019collecting}; the privacy budget $\epsilon$ of the LDP algorithm can also be toggled.
\vspace{-10pt}
\subsection{Other features and specifications}
\begin{figure}[h!]
\centering
\frame{\includegraphics[scale=1]{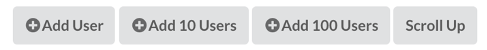}}\vspace{-10pt}
\caption{The add users and scroll up buttons}\vspace{-10pt}
\captionsetup{justification=centering}
\label{fig:addusers}
\end{figure}

An `Add 10 Users' button and an `Add 100 Users' button were added to automatically generate new training data. A scroll up button from the bottom page is added for easier navigation. The buttons are shown in Fig. \ref{fig:addusers}.

Each user's training data is generated via the following equation:
\[ d = -0.55x_1 - 0.82x_2 + 0.07x_3 + 0.95x_4 + 0.31, \]
for linear regression. For logistic regression and SVM it is:
\[ d = 
        \begin{cases}
          1,   &\text{if} -0.55x_1 - 0.82x_2 + 0.07x_3 + 0.95x_4 + 0.31 > 0,\\
          -1,  &\text{otherwise}.\\
        \end{cases}
\]

From these two equations, the ideal weights for the model are $[-0.55, -0.82, 0.07, 0.95, 0.31]$. The closer the model weights are to the ideal weights, the better the model is. Initial model weights and training data are generated with a pseudorandom seed via an extension in the JavaScript Math library.

\subsection{Exp-hamming recovery}
Before the formal definition of the \textit{exp-hamming recovery} is discussed, it is imperative to identify what it means when it is “more difficult” for an aggregator to recover user training data. Consider a user’s training data to be the vector $\vv{x}=(x_{1}, x_{2}, ..., x_{n})$ with $n$ dimensions. Now consider an aggregator’s recovered training data to be the vector $\vv{x_r}=(x_{1r}, x_{2r}, \cdot, x_{nr}$). The natural convention is that it is more difficult to recover $\vv{x}$ if $\vv{x_r}$ is farther from $\vv{x}$. Hence a distance metric is needed; naturally, the higher the value of this metric is, the more difficult it is for an aggregator to recover the user’s training data. Manhattan distance ($\ell1$-norm) is the distance metric used for the definition of the \textit{exp-hamming recovery}.

The \textit{exp-hamming recovery} is designed so that the more (respectively less) difficult it is for the aggregator for recover training data, the lower (respectively higher) the \textit{exp-hamming recovery} should be. The \textit{exp-hamming recovery} is hence formally defined as follows:
\[ E = exp(-k(\|\vv{x} - \vv{x_r}\|_1),\]
where $k$ is a customizable constant (its value is important for accurate interpretation), and $\|\cdotp\|_1$ represents the Manhattan distance.

This metric naturally makes sense because, if $\vv{x}=\vv{x_r}$, there is full recovery of user training data, meaning that $E=1$. If $\|\vv{x}-\vv{x_r}\|_1=\infty$, meaning that $\vv{x}$ and $\vv{x_r}$ are infinitely apart, there is no information gained by the aggregator of what the user training data is like, meaning that $E=0$. One can choose a value of $k$ such that $E<1/e\approx0.368$ should the Manhattan distance $\|\vv{x} - \vv{x_r}\|_1>1/k$. This critical value 0.368 for \textit{exp-hamming recovery} would hence be a good heuristic to whether the aggregator has enough data to be able to make a good guess at the user’s real training data or not. In the demo, the value of $k$ used is 0.5.

\subsection{Privacy budget allocation and training specifications\vspace{-2pt}}
For privacy budget management in each user, the privacy budget is allocated equally to each gradient value. Since there are 5 total values to perturb in the demo, if the privacy budget allocated to each user is $\epsilon$ each of these values are perturbed with privacy budget $\epsilon/5$
For training, for all the machine learning algorithms, stochastic gradient descent proceeds at a constant learning rate of $\alpha=0.01$. A feature to customize this learning rate may be added to a later version of the demo, but the customization of this hyperparameter is not necessary at this stage.

\section{Implementation\vspace{-2pt}}
\begin{figure}[h!]
\centering
\includegraphics[scale=0.9]{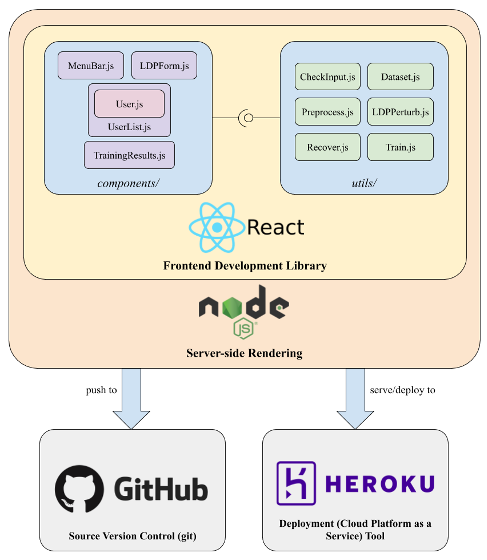}
\caption{Software architecture diagram for ldp-machine-learning}
\captionsetup{justification=centering}
\label{fig:softwarearch}
\end{figure}

\begin{figure}[h!]
\centering
\frame{\includegraphics[scale=0.9]{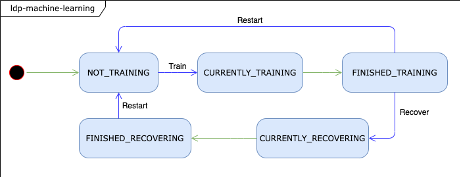}}
\caption{UML State Machine Diagram from ldp-machine-learning}
\captionsetup{justification=centering}
\label{fig:umlstatemachine}
\end{figure}

Fig. \ref{fig:softwarearch} shows the architecture diagram for the web UI, which shows that the site’s UI and main logic operate entirely on the front-end. This is to eliminate the latency from API calls to a backend server, which would slow down the site performance. In addition to that, the UML State Diagram for the UI is shown in Fig. \ref{fig:umlstatemachine}.

\vspace{-5pt}

The development library used for the frontend is React, an open-source JavaScript library. Throughout development, a conscious effort is made to separate the part of the application corresponding to its appearance and to its logic. The appearance is maintained within the project’s ‘components/’ directory, while the site logic lies mainly in the ‘utils/’ directory. The version control system for the project is git and project code is stored on GitHub. Moreover, the demo is deployed and served on Heroku, a cloud platform-as-a-service.

\vspace{-10pt}

\section{Conclusions\vspace{-2pt}}
In this paper, a web GUI is made to illustrate the above results and the power of locally differentially private mechanisms to perturb data. The demo is publicly accessible so potential researchers in the field of differential privacy are able to understand local differential privacy in the context of applying it in machine learning algorithms.

The GUI is easily extensible to other machine learning algorithms and LDP perturbation mechanisms in future work. Moreover, training hyperparameters such as batch size and learning rates can be added to the demo for better training results that nearly match experimental setups with real datasets.

\begin{acks}[font=1pt]
This research was supported by 1) Nanyang Technological University (NTU) Startup Grant, 2) Alibaba-NTU Singapore Joint Research Institute (JRI), 3) Singapore Ministry of Education Academic Research Fund Tier 1 RG128/18, Tier 1 RG115/19, Tier 1 RT07/19, Tier 1 RT01/19, and Tier 2 MOE2019-T2-1-176, 4) NTU-WASP Joint Project, 5) Singapore National Research Foundation (NRF) under its Strategic Capability Research Centres Funding Initiative: Strategic Centre for Research in Privacy-Preserving Technologies \& Systems (SCRIPTS), 6)  Energy Research Institute @NTU (ERIAN), 7) Singapore NRF National Satellite of Excellence, Design Science and Technology for Secure Critical Infrastructure NSoE DeST-SCI2019-0012, 8) AI Singapore (AISG) 100 Experiments (100E) programme, and 9) NTU Project for Large Vertical Take-Off \& Landing (VTOL) Research Platform.
\end{acks}

\Urlmuskip=0mu plus 1mu\relax
\bibliographystyle{ACM-Reference-Format}
\bibliography{references}

\end{document}